\documentstyle[11pt]{article}
\newlength{\breite}
\newcommand{\ab}[1]{\settowidth{\breite}{$#1$} \mbox{\hspace{\breite}}}
\newcommand{\half}{{\textstyle\frac{1}{2}}}

\def\Chi{{\mathop{\kern 2pt\vcenter{\hbox{$\chi $}}\kern2pt}}}

\newcommand{\x}{\begin{equation}\normalsize\begin{array}{rcl}}
\newcommand{\y}{\end{array}\end{equation}}
\newcommand{\be}{\begin{equation}}
\newcommand{\ee}{\end{equation}}
\newcommand{\baym}{\left(\begin{array}{cc}}
\newcommand{\bayv}{\left(\begin{array}{c}}
\newcommand{\eay}{\end{array}\right)}
\newcommand{\beq}{\begin{eqnarray}}
\newcommand{\eeq}{\end{eqnarray}}

\newcommand{\beo}{\begin{eqnarray*}}
\newcommand{\eeo}{\end{eqnarray*}}

\begin{document}

% defines the Feynman--dagger
\def\slash#1{#1 \hskip -0.5em / }

% no \parindent at beginning of paragraphs
\parindent0pt

\thispagestyle{empty}
\begin{titlepage}
\begin{flushright}
%DESY 94--nnn\\
HUB--IEP--94/14 \\
UNIT\"U-THEP-18/1994\\
hep-ph/9409298 \\
September 1994
\end{flushright}
\vspace{0.3cm}
\begin{center}
\Large \bf
The $(0^+,1^+)$ heavy meson multiplet in an
extended NJL model
 \\
\end{center}
\vspace{0.5cm}
\begin{center}
D.\ Ebert\footnotemark[1],
T.\ Feldmann\footnotemark[1], \\
{\sl Institut f\"ur Physik, Humboldt--Universit\"at,\\

 Invalidenstrasse 110, D--10115 Berlin, Germany}
\vspace*{3mm}\\
 R.\ Friedrich$^{\hbox{\footnotesize{2,3,4}}}$,
 H.\ Reinhardt\footnotemark[2], \\
{\sl Institut f\"ur Theoretische Physik, Universit\"at T\"ubingen,\\
 Auf der Morgenstelle 14, D--72076 T\"ubingen, Germany}
\end{center}
\vspace{0.6cm}
\begin{abstract}
\noindent
In this letter we reconsider the previously given description of heavy
mesons within a bosonized extended NJL model that combines heavy quark
and chiral symmetry. In that work the naive gradient expansion of the
quark determinant was used, which satisfactorily works in the light
sector but does not adequately describe the heavy $(0^+,1^+)$ mesons.
By investigating the exact momentum dependence of the quark loop  we
demonstrate that the naive gradient expansion
in the heavy sector is not the right method to treat the
unphysical $q\bar{q}$--thresholds which would be
absent in confining theories.
We propose a modified gradient expansion
which adequately extrapolates from the low--momentum region beyond
threshold. This expansion gives a satisfactory description even of the
$(0^+,1^+)$ heavy mesons  whose masses are significantly above threshold.
\end{abstract}
\vspace{0.3cm}
\setcounter{footnote}{1}
\footnotetext{Supported by
 {\it Deutsche Forschungsgemeinschaft} under contract Eb 139/1--1}
\setcounter{footnote}{2}
\footnotetext{Supported by
 {\it COSY} under contract 41170833}
\setcounter{footnote}{3}
\footnotetext{Supported by a scholarship of the
 {\it Studienstiftung des deutschen Volkes} }
\setcounter{footnote}{4}
\footnotetext{ e--mail: friedric@ptdec5.tphys.physik.uni-tuebingen.de}
\vfill
\end{titlepage}

%%%%%%%%%%%%%%%%%%%%%%%%%%%%%%%%%%%%%%%%%%%%%%%%%%%%%%
% more space between lines
%\baselineskip2em

\setcounter{page}{1}

\section{Introduction}

The observation of new symmetries for infinitely heavy
quarks in QCD together with the formulation of the
Heavy Quark Effective Theory (HQET) \cite{HQET} has
led to a number of theoretical and phenomenological
investigations.
One interesting application is to exploit these
symmetries in the formulation of tractable
quark models \cite{BH93,No93,WE}.

In ref.\ \cite{WE} we have combined chiral symmetry of
light quarks with heavy quark spin and flavor
symmetry in an extension of the Nambu--Jona--Lasinio
(NJL) model. In order to be as close as possible to
former investigations in the light sector \cite{Eb82,Eb86},
we have fixed most of the parameters from light
meson physics.
For evaluating the quark determinant we have
employed a linear gradient expansion.

While our numerical investigations, relating heavy
meson properties and model parameters, have been
successful for the $(0^-,1^-)$ meson doublet of heavy
quark spin symmetry, our approach seems to fail
for the heavier $(0^+,1^+)$ states of opposite
parity. This seems somewhat surprising since
such a problem does not arise in the analogous case of the
$\rho$ or $a_1$ mesons in the light sector.
%\cite{RIPKA}

The aim of this work is to clarify this
problem by a re--investigation of the gradient
expansion in both the heavy and the light meson
sector. The gradient expansion (or equivalently
expansion of quark loops
in external momenta) is usually performed,
in order to avoid unphysical quark thresholds
in NJL--type models without quark confinement.
In fact, assuming such models valid
in the low--momentum region far below the thresholds,
a gradient expansion can afterwards be interpolated to
external momenta of even a few 100 MeV above the
threshold where it still predicts satisfactory results \cite{Eb82,Eb86}.

Nevertheless, the situation for heavy and light
mesons is qualitatively different due to the
different analytical structure of quark loop
expressions involving a heavy quark propagator
$(v \cdot k + i \epsilon)^{-1}$. Therefore,
one should look carefully how to define
the interpolation procedure to larger external momenta which is
necessary to describe the heavier states.

The organization of the letter is as follows:
In section 2 we will give a short review of the
definition of the extended NJL model for heavy quark flavors and of the
result for the heavy meson self--energy which
defines masses and renormalization factors of
heavy mesons \cite{WE}. We show that the
$(0^-,1^-)$ and $(0^+,1^+)$ heavy mesons cannot
be described simultaneously by one set of
parameters.
The next section 3 is devoted to a detailed analysis
of the gradient expansion by evaluating the exact
dependence on external momenta of the pertinent loop diagrams in both
the light and
heavy sector. Our analysis shows that
a rather different way of interpolation is required for the
heavy meson self--energy as compared to the light sector. This
suggests a modified gradient expansion which is valid in both the
light and heavy sector. The modified gradient expansion is shown to lead to
the desired improvement in numerical results.
Finally, some concluding remarks are
given in section 4.

%%%%%%%%%%%%%%%%%%%%%%%%%%%%%%%%%%%%%%%%%%%%%%%%%%%%%%%

\section{Free effective meson lagrangian and lowest--order
         gradient expansion }
\label{model}

In ref.\ \cite{WE} we have presented an
extension of the NJL model which combines
chiral symmetry for light quarks
with heavy quark symmetries for heavy quark fields defined  by
\beq
Q_v(x) = \frac{1 + \slash{v}}{2} \exp (im_Q v\cdot x) Q(x)
\quad .
\eeq
We do not want to give technical details of the
employed bosonization procedure but simply
quote the result for the effective meson lagrangian when all the
quark fields have been integrated out
\beq
{\cal L} & =
& -iN_c {\rm Tr} \ln i\slash{D}
-\frac{1}{4G_1} {\rm tr_F}
 \left[ \Sigma^2 -
 \widehat{m}_0 (\xi \Sigma \xi + \xi^\dagger \Sigma \xi^\dagger)
 \right]
\nonumber\\
&&+\frac{1}{4G_2} {\rm tr_F}
 \left[ (V_\mu - {\cal V}_\mu^\pi)^2
       +(A_\mu - {\cal A}_\mu^\pi)^2
 \right]
\nonumber \\
&& + \frac{1}{2G_3}
 {\rm Tr} \left[ (\overline{H} + \overline{K})\, (H - K) \right]
\quad ,
\label{eff}
\end{eqnarray}
where $N_c = 3$ is the color factor, $\hat{m}_0$ is the current mass
matrix of light quark flavors and $G_1$, $G_2$, $G_3$ are
coupling constants of four--quark interactions.
Furthermore,
\beq
i \slash{D} = i \slash{\partial}
- \Sigma
        + \slash{V} + \slash{A}\gamma_5
        - (\overline{H}+\overline{K})\, (i v \cdot \partial)^{-1}
              (H+ K)
\quad
\label{Dirac}
\eeq
is the Dirac operator for the light quarks, which contains the
several light $(\Sigma , V_\mu , A_\mu )$ and heavy meson fields $(H,K)$.
In the heavy quark limit, heavy meson fields
are organized in spin symmetry doublets
\beq
H &=& P_+
      (i \Phi^5 \gamma_5 + \Phi^\mu \gamma_\mu)
\quad , \qquad v_\mu \Phi^\mu = 0 \quad ,\\
K &=& P_+
      (\Phi  + i \Phi^{5\,\mu} \gamma_\mu \gamma_5)
\quad , \qquad v_\mu \Phi^{5\,\mu} = 0 \quad ,
\eeq
with $\Phi$ being the heavy scalar, $\Phi^5$ the heavy
pseudoscalar, $\Phi^\mu$ the heavy vector and $\Phi^{5\,\mu}$
the heavy axial vector field, respectively.
$V$, $A$ denote the light vector and axial--vector fields.
For the light octet of Goldstone bosons we use the common
non--linear representation
$\xi = \exp (i \pi / F)$
where $\pi = \pi^a \lambda_F^a/2$ and $F$ is the bare
decay constant. Vector and axial--vector expressions are build
via
$ {\cal V}_\mu^\pi = i/2 (\xi \partial_\mu \xi^\dagger +
             \xi^\dagger\partial_\mu \xi)$,
$ {\cal A}_\mu^\pi = i/2 (\xi \partial_\mu \xi^\dagger -
             \xi^\dagger\partial_\mu \xi)$.
Finally, the light scalar field $\Sigma$ achieves a non--vanishing
vacuum expectation value indicating the spontaneously breaking of
chiral symmetry.

 From (\ref{eff}) we have derived self--energy expressions for
mesons
as well as interaction terms between heavy and light mesons.
This is achieved by evaluating ${\rm Tr} \ln i \slash{D}$ in terms
of quark loops that are regularized by Schwinger's gauge
invariant proper--time method.
Note that due to chiral and heavy quark symmetry, the heavy
sector is controlled by only one coupling constant $G_3$ for
both $H$ and $K$ fields, while all other parameters are fixed
from the light sector.

%%%%%%%%%%%%%%%%%%%%%%%%%%%%%%%%%%%%%%%%%%%%%%%%%%%%%%%

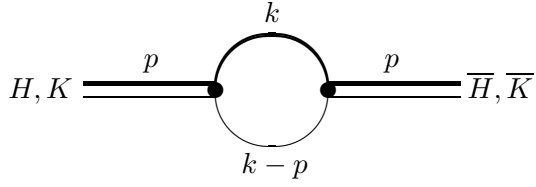
\begin{figure}
\begin{center}
\unitlength1cm
\begin{picture}(5,3)
\thicklines
\put(0,1.6){\line(1,0){1.75}}
\put(0,1.57){\line(1,0){1.75}}
  \put(0.8,1.8){\mbox{$p$}}
\put(3.25,1.6){\line(1,0){1.75}}
\put(3.25,1.57){\line(1,0){1.75}}
  \put(4,1.8){\mbox{$p$}}
\put(2.5,1.5){\oval(1.5,1.5)[t]}
\put(2.5,1.47){\oval(1.5,1.5)[t]}
  \put(2.4,2.4){\mbox{$k$}}
\thinlines
\put(2.5,1.5){\oval(1.5,1.5)[b]}
  \put(2.1,0.4){\mbox{$k-p$}}
\put(0,1.4){\line(1,0){1.75}}
\put(3.25,1.4){\line(1,0){1.75}}
\put(1.75,1.5){\circle*{0.2}}
\put(3.25,1.5){\circle*{0.2}}
 \put(-1,1.4){\mbox{$H,K$}}
 \put(5.1,1.4){\mbox{$\overline{H},\overline{K}$}}

\end{picture}
\end{center}
\caption{Self--energy diagram for heavy meson fields $H,K$.}
\label{dia1}
\end{figure}

The self--energy term for heavy mesons
$\Pi_{H,K}$ as a function of
the external momentum $(v\cdot p)$ is
represented in figure \ref{dia1} and calculated as
\begin{eqnarray}
&& - {\rm tr_D}
  \left[ \overline{H}^i \Pi_H^i (v\cdot p) H^i \right]
 +  {\rm tr_D}
  \left[ \overline{K}^i \Pi_K^i (v\cdot p) K^i \right]
\nonumber \\
&=&
 i N_c \int^{reg} \frac{d^4k}{(2\pi)^4}
  \frac{ {\rm tr_D} \left[ (\slash{k} -\slash{p} + m^i)\,
                  (\overline{H}^i + \overline{K}^i)\, (H^i + K^i)
  \right]}
  {\left( (k-p)^2-(m^i)^2 \right)\left(v\cdot k + i\epsilon\right)}
\quad ,
\label{se}
\end{eqnarray}
where $ i=u,d,s $ is a light flavor index.

Remember that the heavy $(Q_v\bar{q})$--meson lagrangian in
the heavy quark limit is given by\footnote{
Note that the heavy meson field has the mass
dimension $3/2$.}
\beq
 {\cal L}_0^{heavy} & = & - {\rm tr_D}
            \left[\overline{H}^i (iv\cdot \partial - \Delta
	    M_H) H^i
            \right]
\nonumber \\
          &&     +  {\rm tr_D}
            \left[ \overline{K}^i (iv\cdot \partial - \Delta
	    M_K) K^i
             \right]
\quad ,
\label{prop}
\eeq
where $\Delta M_{H,K}^i = M_{H,K}^i - m_Q$, with $M_{H,K}^i$ being the
heavy meson mass.
Let us compare the inverse
propagator term $(v\cdot p - \Delta M_{H,K}^i)$ resulting
from (\ref{prop}) with the self--energy expression
$\Pi_{H,K}^i$ of (\ref{se}) and the term
$ \sim 1/G_3 $ in (\ref{eff}). First observe that $\Pi_{H,K}^i$ is a
function of $v\cdot p$ only.
Expanding $\Pi_{H,K}^i$ around the
on--shell value $v \cdot p = \Delta M_{H,K}^i$, one obtains
\beq
\Pi_{H,K}^i(v\cdot p) = \Pi_{{H,K}}^i(\Delta M_{H,K}) +
\Pi_{{H,K}}^{'\,i}(\Delta M_{H,K}) \, \left(v\cdot p - \Delta M_{H,K}\right)
+ O((v\cdot p - \Delta M_{H,K})^2)
\quad   .
\label{self}
\eeq
With this expansion one reads off from (\ref{se}) that the meson mass
is determined by
\beq
I_3^i(\Delta M_{H,K}^i) \left(\Delta M_{H,K}^i \pm m^i\right) + I^i_1
-\frac{1}{2G_3} = 0\quad
\label{selfcon}
\eeq
and the $Z$--factors for the necessary field renormalization
$H^i,K^i \to (Z^i_{H,K})^{-1/2} H^i,K^i$ are given by
\beq
Z^i_{H,K}(\Delta M_{H,K}^i) = \left(I_3^i(\Delta M_{H,K}^i) +
2 I_2^i(\Delta M_{H,K}^i)\,
(\Delta M_{H,K}^i \pm m^i)  \right)^{-1}\quad ,
\label{Zfac}
\eeq
where $I_1^i$, $I_3^i$ are loop integrals defined by
\beq
I_1^i &=& \frac{iN_c}{16\pi^4} \int^{reg}
          \frac{d^4k}{k^2 - (m^i)^2} \nonumber \\
    &=& \frac{N_c}{16\pi^2} (m^i)^2 \Gamma(-1,(m^i)^2/\Lambda^2)
\quad ,\\
I_3^i\left( v\cdot p\right) &=& - \frac{iN_c}{16\pi^4} \int^{reg}d^4k
            \frac{1}{(k^2 - (m^i)^2)(v\cdot k +v\cdot p + i\epsilon)}
\nonumber \\
    &=& \frac{N_c}{16\pi^2}\left\{ (v\cdot p)
    \int_0^1 dx \left(1-x\right)^{-1/2}
\Gamma\left(0,\frac{(m^i)^2-x(v\cdot p)^2}{\Lambda^2}\right)
\right.\nonumber\\& &\ab{\frac{N_c}{16\pi^2}}\left.
+\sqrt{\pi}\sqrt{(m^i)^2-(v\cdot p)^2}
\Gamma\left(-\half ,\frac{(m^i)^2-(v\cdot p)^2}{\Lambda^2 } \right)
         \right\}\quad ,
\label{i3}
\eeq
$I_2^i (v\cdot p) = 1/2 \, d I_3^i (v \cdot p) / d (v\cdot p)$
and $\Gamma(\alpha,x)$ is the incomplete gamma function.

As a consequence of the heavy flavor symmetry the solutions of
(\ref{selfcon}) do not depend on the heavy quark mass. But they do
depend on the light quark constituent masses and so are influenced by
the explicit and spontaneous chiral symmetry breaking.
For that reason we have
supplied the mass shifts $\Delta M_{H,K}^i$ with the index $i$ of the
light quark flavors.

It is obvious that
for $v\cdot p > m^i$ the quark loop produces an imaginary part due
to the unphysical quark--antiquark threshold.
Clearly, one must neglect this imaginary part of the meson self energy,
since it would yield a nonvanishing decay width
$\Gamma(H^i,K^i \to \bar{q}^i + Q)$
forbidden by confinement.
Its formal appearance is obviously related to the fact that
confinement is not taken into account in our model explicitely.
The simplest way to avoid these unphysical two--quark thresholds is
to expand $\Pi_{H,K}^i$ to first order around
$(v\cdot p)=0$
in analogy to the gradient expansion of the light meson sector, where
one expands to
first order around $p^2=0$ \cite{Eb82,Eb86}.
This lowest--order expansion in $(v\cdot p)$
has been performed in ref.\ \cite{WE}, leading to
\beq
\Delta M_{H,K}^i\left(I_3^i(0)\pm 2 m^i I_2^i(0) \right)\pm
m^i I_3^i(0)+I_1^i - \frac{1}{2G_3} = 0\quad ,
\label{gradm} \\
Z_{H,K} = \left(I_3^i(0)\pm 2 m^i I_2^i(0)   \right)^{-1}\quad .
\label{gradz}
\eeq
Note that eq.\ (\ref{gradm}) is linear in $\Delta M$, and the
Z--factors in eq.\ (\ref{gradz}) are independent of $\Delta M$.

Let us also recall the numerical results following from our
simple gradient expansion (\ref{gradm},\ref{gradz}).
While the values of the several mass differences
$\Delta M_{H,K}^i$
are to be compared with experimental values,
the $Z_{H,K}^i$--factors enter into the decay constants
$f_{H,K}$ of heavy mesons defined by
\beq
    \langle 0| \bar{q} \gamma_\mu (1-\gamma_5) Q_v|H_v(0^-)\rangle
  &=& i f_{H} M_{H} v_\mu \nonumber \quad ,\\
   \langle 0| \bar{q} \gamma_\mu (1-\gamma_5) Q_v|K_v(0^+)\rangle
  &=&  - f_{K} M_{K} v_\mu
\quad ,
\eeq
via the equation \cite{WE}
\beq
  f_{H,K} \sqrt{M_{H,K}} = \frac{\sqrt{Z_{H,K}}}{G_3}
\quad .
\label{law}
\eeq
The light parameters $m^{u,d} = 300$ MeV,
$m^s = 510$ MeV, $\Lambda = 1.25$ GeV,
entering (\ref{gradm},\ref{gradz}) are
fixed from light meson properties, whereas the
heavy--light quark coupling constant
$G_3$ has to be adjusted from heavy meson physics.

As table \ref{tableG3} shows, it is
impossible to find a consistent value of $G_3$ that fits
both, the masses of heavy $H$ and $K$ mesons simultaneously.
While a value of $G_3 = 8.7$ GeV$^{-2}$ gives realistic
results for masses and decay constants of the members of
the $H(0^-,1^-)$ multiplet \cite{WE},
the masses and Z--factors for
the parity conjugate mesons $K$ come out simply too large.
This is in contrast to the light flavor sector where both vector and axial
vector mesons are adequately reproduced. Obviously something must go
wrong in the gradient expansion in the heavy sector.
We therefore abandon the naive gradient expansion and re--investigate
the exact expressions (\ref{selfcon},\ref{Zfac}).
This will lead us to a modified gradient expansion which overcomes the
shortcomings of the naive gradient expansion in the heavy sector.
It is important to note that such a modification of the gradient
expansion should not be understood as a better approximation
(recall that we even know the exact result) but rather as a
heuristic method to
extract physical information from quark loop calculations
without including unphysical
threshold effects. We hope that the following discussion
will illuminate these ideas.

\begin{table}
\begin{center}
\begin{tabular}{|c||c|c|c|c|||c||c|}
\hline
$G_3 [ \mbox{GeV}^{-2} ] $ &5&10&15&20&Exp.& \\
\hline\hline
$\Delta M_H^s-\Delta M_H^{u,d} \left[ \mbox{MeV}\right]$ &
240 & 80 & 20 & $<$0 & 100 &$M_{D_s} - M_D\left[ \mbox{MeV}\right]$ \\
\hline
$ \Delta M_K^{u,d}-\Delta M_H^{u,d}\left[ \mbox{MeV}\right]$ &
2710 & 1450 & 1030 & 820 & 410 & $M_{D_1} - M_D\left[ \mbox{MeV}\right]$\\
\hline
$\Delta M_K^{s}-\Delta M_H^{u,d}\left[ \mbox{MeV}\right]$ &
9020 & 4780 & 3370 & 2670 & 530 & $M_{D_{s1}} - M_D\left[ \mbox{MeV}\right]$ \\
\hline
\end{tabular}
\end{center}
\caption{Differences of the mass shifts as a function of the
heavy--light quark coupling
constant $G_3$. The column "Exp." contains the experimental values.}
\label{tableG3}
\end{table}

\section{Modified gradient expansion}

Let us now start to investigate the exact equations
(\ref{selfcon},\ref{Zfac}). Since we expect the mass
shifts $\Delta M_K^i$ to be larger than $m^i$ we have
to handle the imaginary part of the integral
(\ref{i3}) above the threshold.
For illustration let us drop the imaginary part
and show the real part of eq.\ (\ref{selfcon}). In
figure 2 we have plotted, for convenience,
 $G_3^{-1}$ as
a function of $\Delta M_{H,K}^u$ resulting from the
exact solution (\ref{selfcon}) and from the simple
gradient expansion (\ref{gradm}), respectively.
(Heavy mesons with strangeness show a completely similar behaviour.)
\input epsf
%
%begin figure 2
%
\begin{figure}
\epsfbox{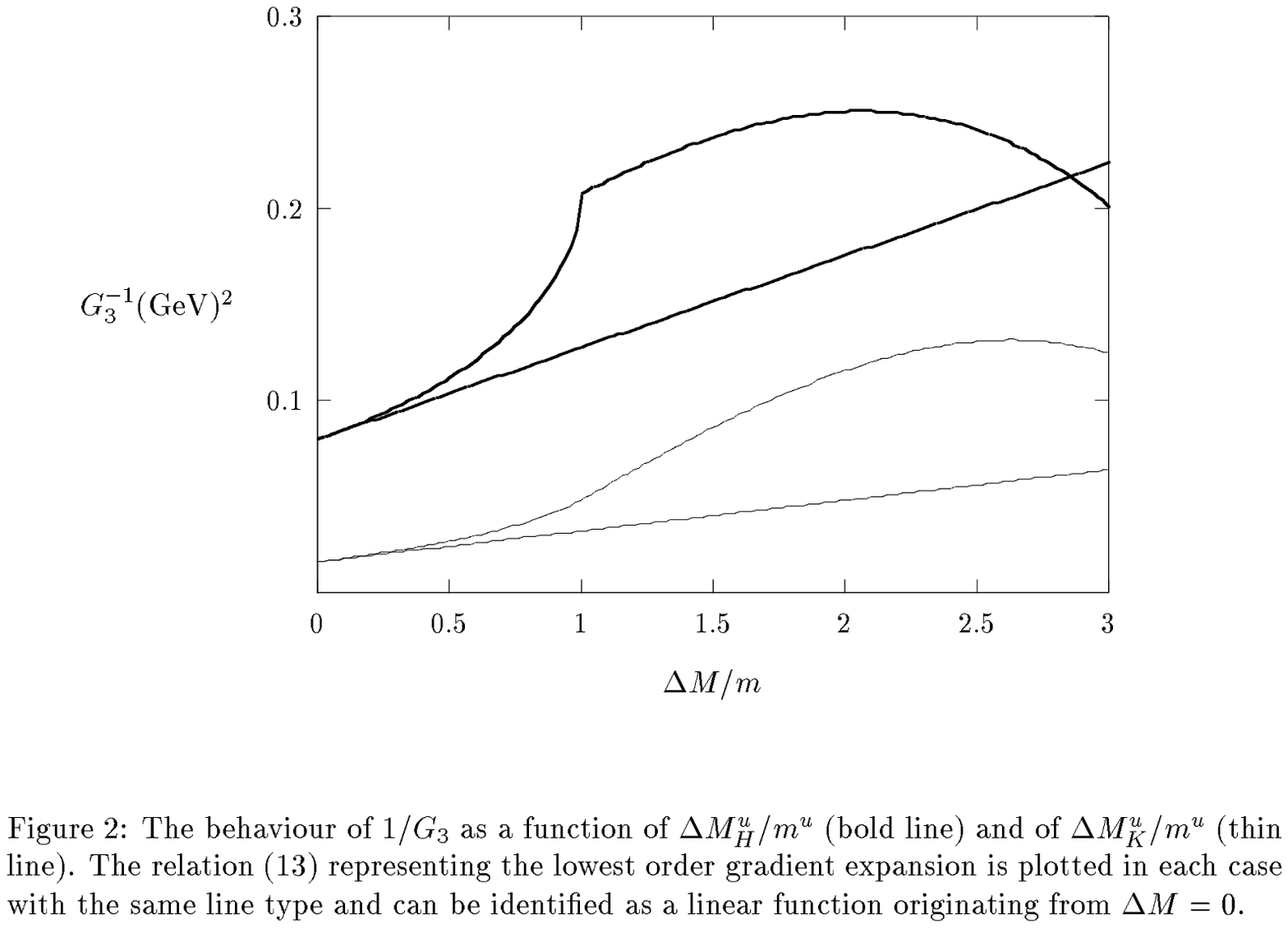}
%\label{figure2}
\end{figure}
%
%end Figure 2
%
Note that already for values below the treshold $\Delta M =m$
the linear approximation seems problematic.
Instead, $G_3^{-1}$ as a function of $\Delta M$ seems to
start already with a significant curvature. Keeping more of
the quadratic behaviour at $\Delta M = 0$ would immediately
decrease $\Delta M_K$ for a given value of $G_3$
considerably, whereas $\Delta M_H$ would not such
dramatically change.

For clarification let us compare the renormalization
factors $Z_{H,K}^i$ in the exact formula (\ref{Zfac})
with the simple gradient expansion (\ref{gradz}).
As an example we plotted $(Z_H^u)^{-1}$ in figure 3.
%
%begin figure 3
%
\begin{figure}
\epsfbox{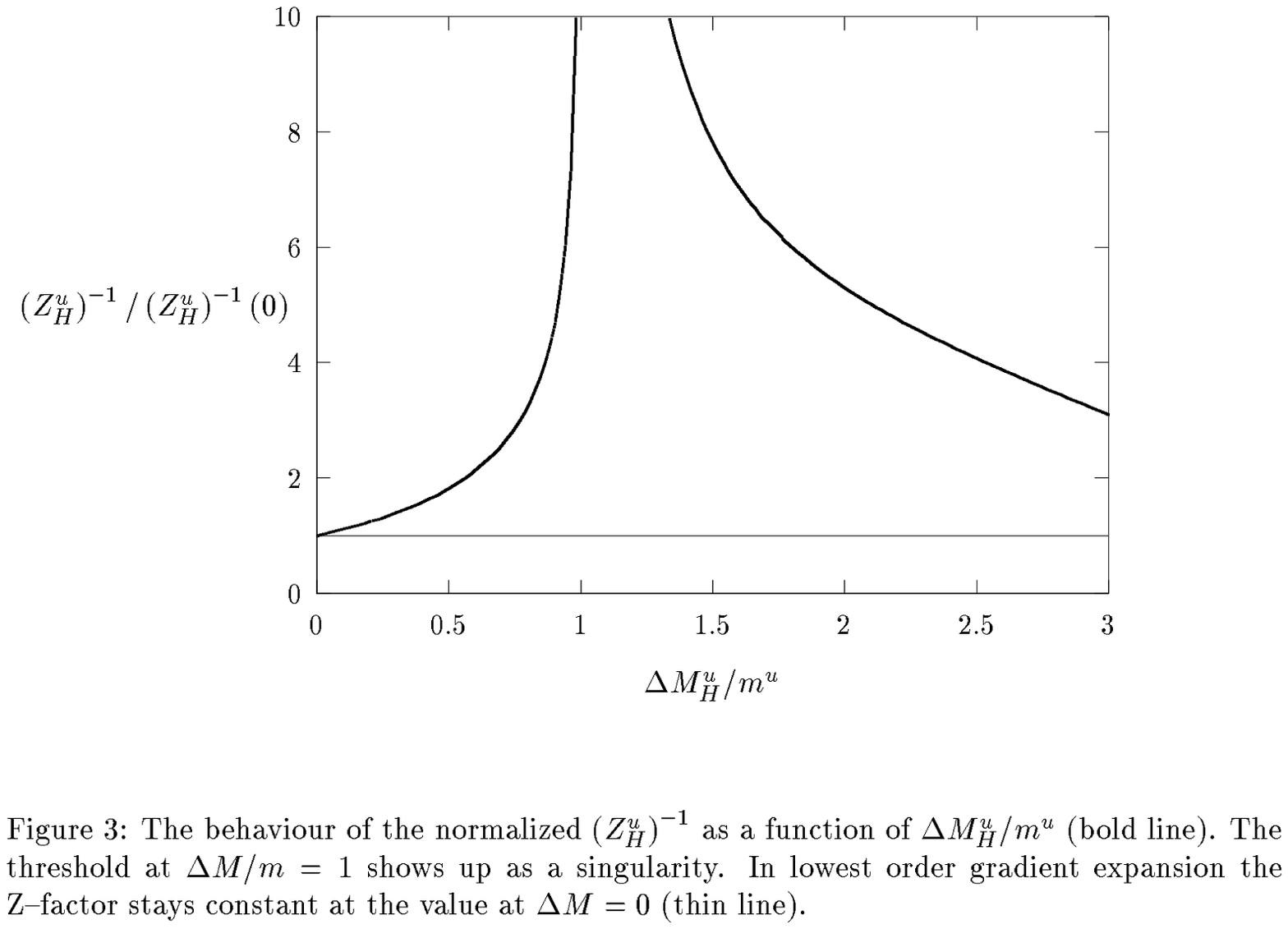}
%\label{figure3}
\end{figure}
%
%end Figure 3
%
Once again the simple gradient expansion cannot be viewed
as a good low--energy approximation. This becomes even more
appearent if we consider the analogous situation for the
light sector. In figure 4 we plotted the
real part of an exact calculation of the
renormalization factor for the $\rho$ meson given by
\beq
Z_\rho^{-1} (p^2) =
\frac{N_c}{16 \pi^2} \int_0^1 dx 4 x (1-x)
\Gamma(0, \frac{m^2 - x (1-x) p^2}{\Lambda^2})
\quad .
\eeq
Here the usual gradient expansion is indeed justified as
a definite interpolation from the low energy region to
regions above the threshold.
%
%begin figure 4
%
\begin{figure}
\epsfbox{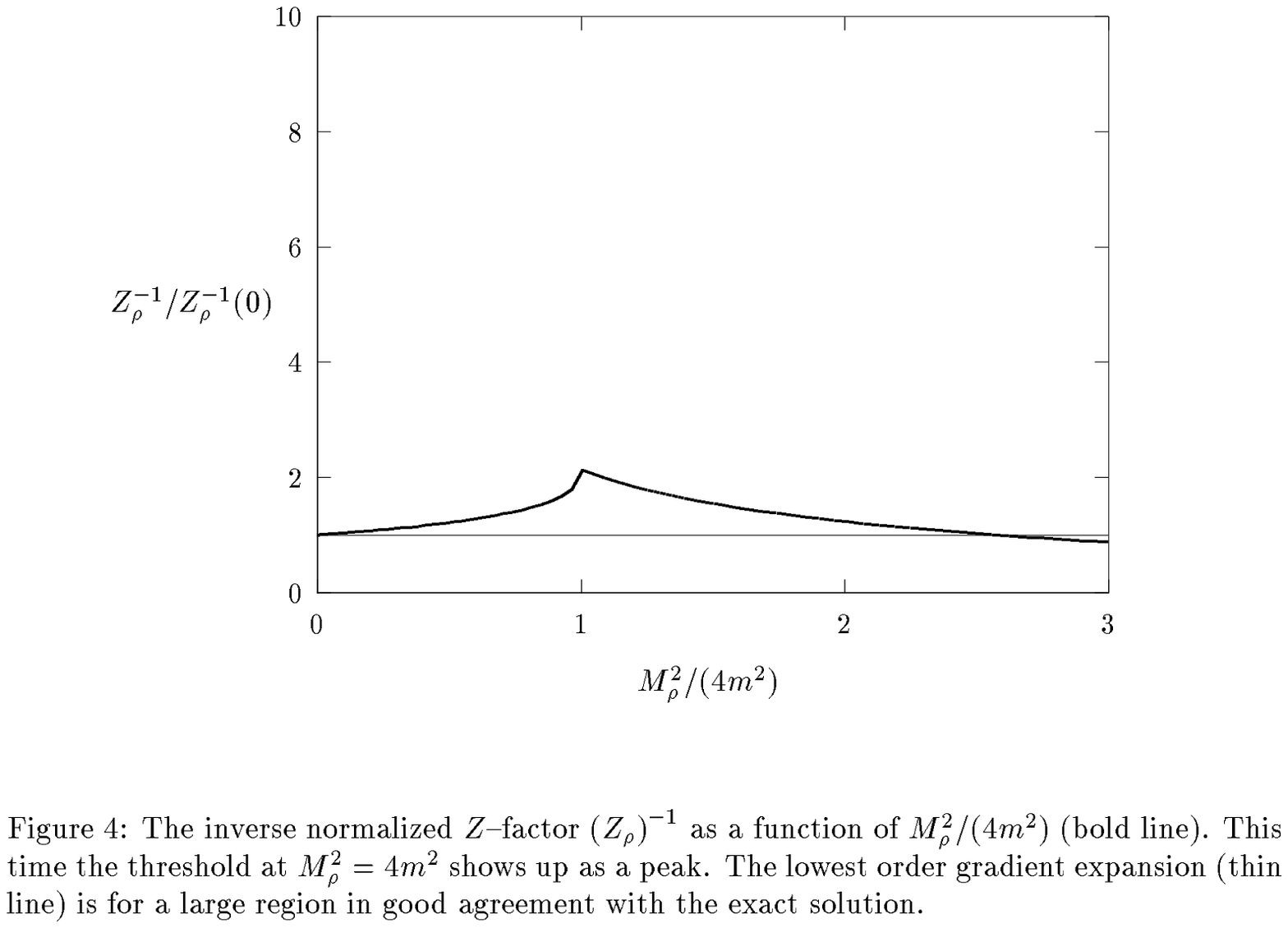}
%\label{figure4}
\end{figure}
%
%end Figure 4
%

Let us now try to understand analytically which terms are
responsible for the deviation
from  the linear behaviour of eq.\ (\ref{selfcon})
(respectively from the constant behaviour of the Z--factors)
in the low--energy region.
The threshold
presents itself in the combination $m^2 - (x) \Delta M^2$
as one sees from the integral $I_3 ((v\cdot p) = \Delta M)$ in
eq.\ (\ref{i3}).
For $\Delta M > m$ the incomplete gamma function and the square root
with negative argument will produce imaginary parts. In the
proper--time formalism this is the usual way how thresholds
arise. Since the low--energy region is defined as
$\Delta M \ll m$ we are on the safe side if we neglect
$\Delta M^2$ compared to $m^2$ in the integral expressions,
performing the replacement
\beq
I_3(\Delta M)&\rightarrow& \frac{N_c}{16\pi^2}\left\{
2\Delta M \Gamma\left(0,\frac{m^2}{\Lambda^2}\right)
+\sqrt{\pi}m
\Gamma\left(-\half ,\frac{m^2}{\Lambda^2 } \right)
         \right\}\nonumber \\
         &=& I_3(0) + 2\Delta M I_2(0)
         \quad , \\
I_2(\Delta M) &\to& I_2(0) \quad .
\eeq
Indeed such a strategy would reproduce a constant Z--factor
for the $\rho$ meson. Furthermore, for heavy mesons the integral
$I_3$ is multiplied by $(\Delta M \pm m)$, and this will
generate a quadratic term (in $\Delta M$) in the self--energy. For light
mesons such an additional pre--factor is missing.
Note that our so defined modified
gradient expansion has not to be viewed as an ordinary
Taylor expansion of the self--energy.
The crucial point of this investigation is that it would mean no
improvement to calculate any higher order terms in the gradient expansion
since we would get more and more unphysical effects from the threshold.
Instead, our method represents a
compromise for extracting relevant physics from the low energy region
and neglecting unphysical imaginary parts and threshold effects
leading to the emission of free quarks.

One could try to treat the thresholds in the quark
loop integrals seriously and
simply dropping imaginary parts. For the light (axial) vector mesons this would
(accidently) make not much difference. However for heavy mesons
such a strategy gets no numerical support and seems unphysical
at all.

With this understanding we can dare to extrapolate from the region
below threshold to mass shifts $\Delta M$ above
threshold. Doing so we hope that chiral symmetry dominates the physics
over a large region, at least so far that also the $(0^+,1^+)$
multiplet can be described by our modified gradient expansion.
Equations (\ref{selfcon}), (\ref{Zfac}) are then to be replaced by
\beq
\left(I_3^i(0)+ 2I_2^i(0)\Delta M_{H,K}^i \right)
\left(\Delta M_{H,K}^i\pm m^i\right)
 + I_1^i - \frac{1}{2G_3} = 0
\quad ,
\label{mgrad2}\\
\left( Z_{H,K}^i\right)^{-1} = I_3^i(0) +2 \left(2\Delta M_{H,K}^i \pm
m^i \right)I_2^i(0)
\label{zgard2}
\quad .
\eeq
Figure 5 shows the behaviour of $G_3^{-1}$ in the
modified gradient expansion (\ref{mgrad2}) to be compared
with figure 2.
%
%begin figure 5
%
\begin{figure}
\epsfbox{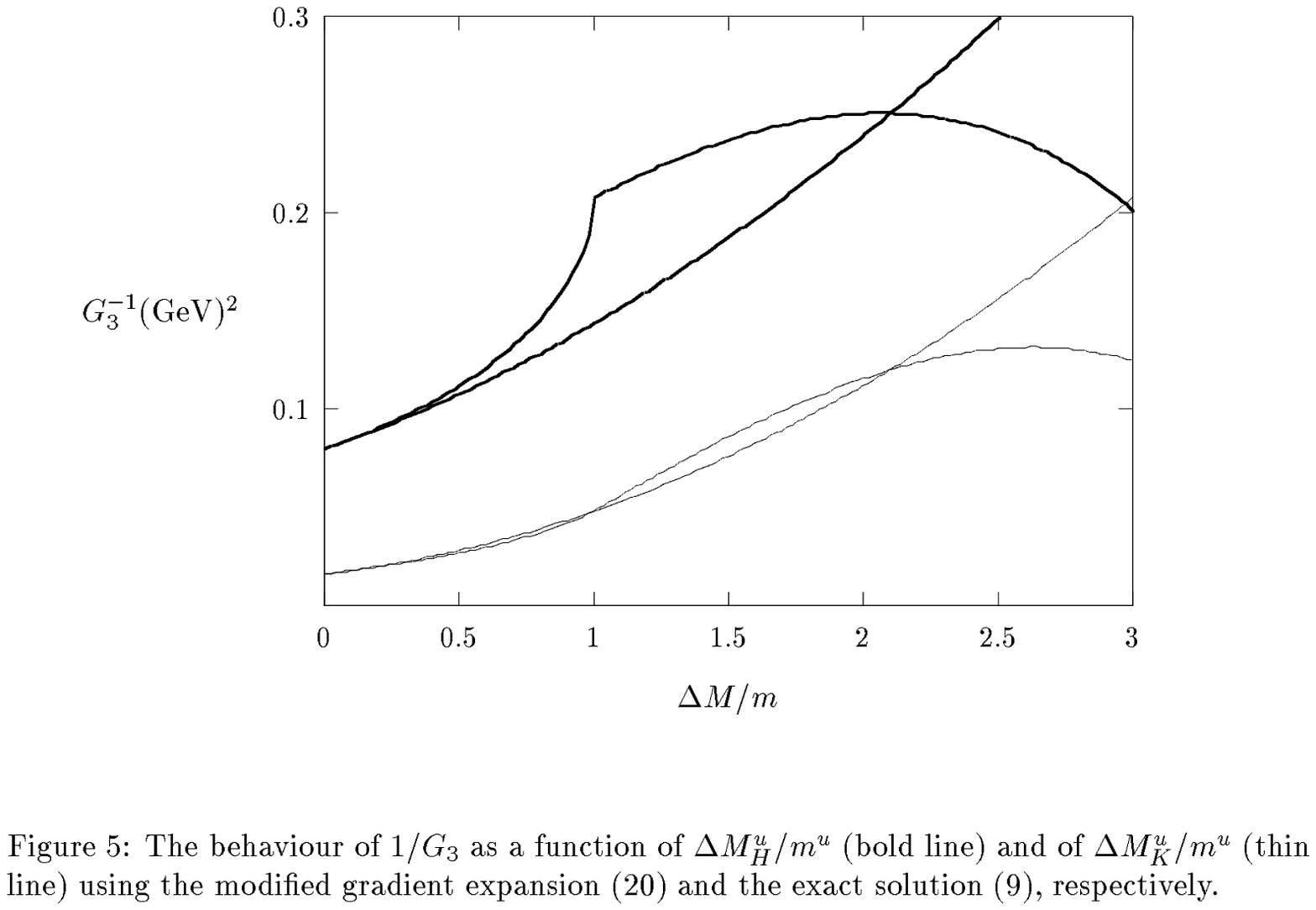}
%\label{fig5}
\end{figure}
%
%end Figure 5
%

In table \ref{tableall} we present the numerical results
following from this
modified gradient expansion.
\begin{table}[hbt]
\begin{center}
\begin{tabular}{|c||c|c|c|c|c|||c||c|}
\hline
$G_3 [ \mbox{GeV}^{-2} ] $ &3&5&7&9&11& "Exp."& \\
\hline\hline
$m_c \left[ \mbox{MeV}\right]$ &
1070 &1410 & 1600 & 1740 & 1840 &1550 & $M_{J/\psi}/2 \left[ \mbox{MeV}\right]$
\\
\hline
$m_b \left[ \mbox{MeV}\right]$ &
4470 &4810 & 5000 & 5140 & 5240 & 4730 & $M_\Upsilon/2 \left[
\mbox{MeV}\right]$ \\
\hline
$\Delta M_H^s-\Delta M_H^{u,d} \left[ \mbox{MeV}\right]$ &
250 &160 & 110 & 80 & 50 & 100 &$M_{D_s} - M_D\left[ \mbox{MeV}\right]$ \\
\hline
$ \Delta M_K^{u,d}-\Delta M_H^{u,d}\left[ \mbox{MeV}\right]$ &
370 & 390 & 410 & 430 & 450 & 410 & $M_{D_1} - M_D\left[ \mbox{MeV}\right]$\\
\hline
$\Delta M_K^{s}-\Delta M_H^{u,d}\left[ \mbox{MeV}\right]$ &
870 & 800 & 780 & 780 & 780 & 530 & $M_{D_{s1}} - M_D\left[ \mbox{MeV}\right]$
\\
\hline
$f_B\left[ \mbox{MeV}\right]$ &
300 &210 & 170 & 150 & 130 & 180 &  \cite{Latt,Sum} \\
\hline
$f_H^s/f_H^u$ &
1.13 & 1.13 & 1.12 & 1.11 &1.09 & 1.1--1.2& \cite{Latt,Sum} \\
\hline
\end{tabular}
\end{center}
\caption[]{Here we show different observables as a function of the coupling
constant $G_3$ calculated with the modified gradient expansion. The column
"Exp." contains the experimental values or in
case of the heavy constituent quark masses some plausible estimates.
In the last column we give the way of determination. (For
the electroweak decay constants we have given the values calculated
from the lattice and QCD sum rules.)
}
\label{tableall}
\end{table}
Note that heavy quark masses are extracted from
$m_{b,c} = M_{B,D} - \Delta M_H^u$. For the calculation
of heavy meson decay constants we have used the experimental
values of meson masses \cite{PD} if available, or have made
use of the relation $M_{B_x} - M_B = M_{D_x} - M_D$ which
is true in the heavy quark limit ($x = \{ s , 1 , s1\}$).

Indeed we observe a drastic improvement compared to the
results in table \ref{tableG3}. For a value $G_3 \approx 7$ GeV$^{-2}$
a simultaneous fit of nearly all heavy meson properties is
possible. Only for the most massive $(B_{s1},D_{s1})$ states,
masses are still predicted slightly too high\footnote{
This indicates that a simple local interaction
is perhaps not sufficient to describe a heavy quark weakly bound
via a p--wave with a not so light strange quark.}.

With our analysis we are now able to predict the decay
constant of several B--mesons (Note that $f_D$ is
assumed to get large $1/m_c$ corrections). Within a physical
acceptable range of
\beq
6.5 \mbox{ GeV}^{-2} < G_3 < 7.5 \mbox{ GeV}^{-2}
\eeq
we obtain
\beq
160 \mbox{ GeV} < &f_B   &  < 180 \mbox{ GeV} \\
180 \mbox{ GeV} < &f_{B_s}& < 200 \mbox{ GeV}\\
150 \mbox{ GeV} < &f_{B_1}& < 160 \mbox{ GeV}\\
155 \mbox{ GeV} < &f_{B_{s1}}& < 175 \mbox{ GeV}
\quad .
\eeq

%%%%%%%%%%%%%%%%%%%%%%%%%%%%%%%%%%%%%%%%%%%%%%%%%%%%%%%%%%%%%%%%%%%%%

\section{Conclusions}

In this paper we have studied in detail the treatment of
quark loop diagrams, arising from the calculation of
the quark determinant in the recently proposed extended
NJL model \cite{WE}.

Usually, the problems originating from the lack of confinement in quark models
which
leads to unphysical quark thresholds, are overcome
by an expansion in external momenta. The lowest--order
gradient
expansion works successfully for composite light mesons.
However, in the heavy meson sector there is phenomenological
evidence that the gradient expansion has to be improved,
which has been discussed in section 2.

We have shown that this fact is due to an essentially different
analytical structure of the loop integrals under concern, coming
from the special form of the heavy quark propagator
$(v\cdot k + i \epsilon)^{-1}$. While the renormalization
factor $Z_\rho$ for $\rho$ mesons stays nearly constant
below the threshold, this is not the case
for the corresponding expressions $Z_{H,K}$ of heavy mesons.
We have convinced ourselves that this
additional momentum dependence is not due to threshold
effects and should be included. We stress again that
our modified procedure should not be viewed as an
ordinary Taylor expansion that would never be able
to interpolate beyond the threshold.

Most importantly, with this improved gradient expansion
heavy meson properties of both, the $(0^-,1^-)$ and
the $(0^+,1^+)$ spin symmetry doublet
can now be described
simultaneously with a quark--coupling constant
$G_3 \approx 7$ GeV$^{-2}$.
With this value we are in a position to estimate several
weak decay constants of B mesons.

In summary, it has been shown that the evaluation of
quark loop diagrams with heavy quark propagators in
unconfined models demands a modification of the gradient
expansion that has been used in the light sector.
A phenomenologically successful procedure seems to consist in neglecting
unphysical two--quark threshold effects in a definite
way and keeping all remaining momentum dependence.
Phenomenologically, this leads to realistic results
for heavy meson masses and decay constants within our
extended NJL model with chiral and heavy quark symmetries.

\section*{Acknowledgements}

Two of the authors (D.E.\ and T.F.) thank
Prof.\ D.\ V.\ Shirkov and the
colleagues at the Laboratory of Theoretical
Physics of JINR Dubna for the  hospitality
during their stay in August 1994.

%%%%%%%%%%%%%%%%%%%%%%%%%%%%%%%%%%%%%%%%%%%%%%%%%%%

\end{document}